# Enhanced Biogas Production via Anaerobic Co-Digestion of Slaughterhouse and Food Waste Using Ferric Oxide as a Sustainable Conductive Material

Michelle C. Almendrala *,a, Kyle Adrienne T. Valenzuela[a], Steffany Marie Niña B. Santos[a], Louise Grace S. Avena-Ardeta[a]

[a] *School of Chemical, Biological, and Materials Engineering and Sciences. Mapúa University,*

*Intramuros, Manila, Philippines*

\* Corresponding author.
E-mail address: mcalmendrala@mapua.edu.ph, katvalenzuela@mymail.mapua.edu.ph,

smnbsantos@mymail.mapua.edu.ph, lgsavena@mapua.edu.ph

## ABSTRACT

The anaerobic co-digestion of slaughterhouse wastewater (SHWW) and food waste (FW) offers a sustainable approach to waste treatment and biogas production. However, limited literature was found on the study of $Fe_2O_3$ as conductive material in co-digestion of the two substrates. This study evaluates the effect of $Fe_2O_3$ on biogas yield, organic matter removal, and kinetics of anaerobic co-digestion. Five batch tests were performed — four with varying $Fe_2O_3$ doses and one control. Results showed that $Fe_2O_3$ significantly enhanced total solids (TS) and volatile solids (VS) reduction. The reactor with 0.5 g $Fe_2O_3$ per 800 mL working volume achieved the highest TS and VS reduction, corresponding to the maximum methane yield of 9878.95 L $CH_4$/kg VS. At this optimal dosage, biogas production increased by 81% compared to the control. However, further increases in $Fe_2O_3$ above the optimal dosage concentration decreases biogas yield, indicating a threshold beyond which inhibitory effects occur. In addition, at this optimal dosage, reduction in BOD and COD was observed due to enhanced microbial activity. Furthermore, $Fe_2O_3$ stabilizes anaerobic digestion by mitigating inhibitory compounds and promoting direct interspecies electron transfer (DIET), leading to improved methane yield. Kinetic modeling using the Logistic Function accurately predicted methane production trends, demonstrating its potential for industrial-scale application. Overall, the study confirms that $Fe_2O_3$ at an optimal dose significantly enhance biogas yield and system performance during the anaerobic co-digestion.

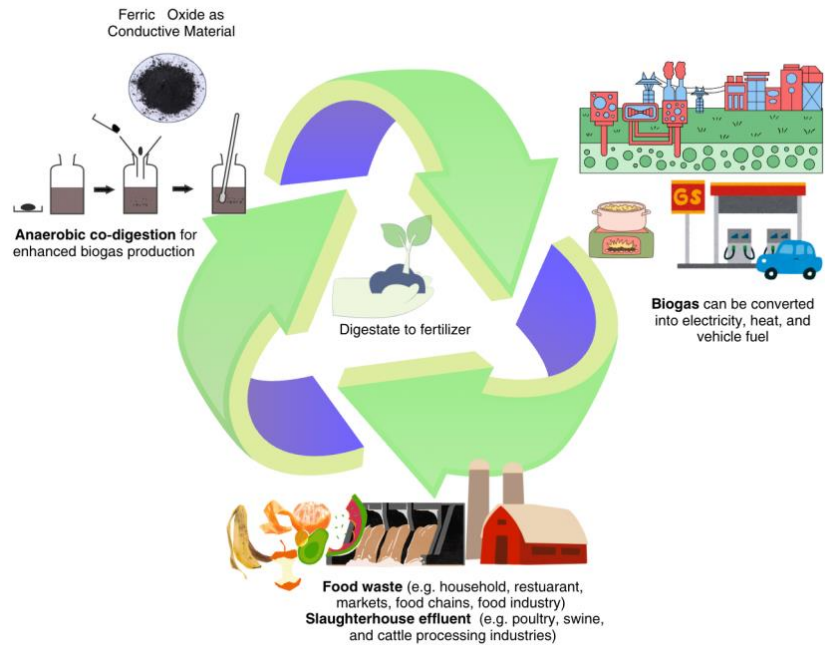

**Highlights**

- By 2025, the Philippines' MSW will reach 28,000 tons daily; 29% of wastewater is from agriculture.
- Anaerobic co-digestion with ferric oxide can increase biogas production and solve waste management problem.
- $Fe_2O_3$ increased rate biogas production through
- improves TS/VS reduction, increases biogas, and stabilizes anaerobic digestion.
- Moderate $Fe_2O_3$ maximizes biogas yield, but higher doses deplete organic matter faster.

**1. Introduction**

The global shift towards renewable energy is needed to address the pressing challenges of climate change, global warming, energy security, and sustainable development. Among renewable energy options, biogas stands out as a particularly valuable resource because it not only provides a sustainable and clean energy source but addresses waste management issues as well. In the Philippines, the MSW, which amounts to 22% of food waste (FW) generated in urban areas, is projected to reach 28,000 tons daily in 2025 [1]. Similarly, 29% of wastewater in the country came from agricultural and livestock industries [2]. Hence, to mitigate the possible negative impact of these wastes on the environment and humans, it is necessary to have a series of treatments, and one viable option is to use these as substrates in the anaerobic digestion process.

Emerging studies highlight the role of conductive materials, such as $Fe_2O_3$, in enhancing anaerobic digestion (AD) performance. Very limited literature is found in the anaerobic co-digestion of slaughterhouse wastewater and food waste, particularly with the addition of ferric oxide. This study aims to investigates the potential of Fe₂O₃ to enhance biogas production and digestion performance during the anaerobic co-digestion of SHWW and FW. In addition, the study will (a) evaluate the effects of Fe₂O₃ on methane yield during co-digestion; (b) identify the physicochemical properties of digestate that correlate with optimal biogas production (c) determine the optimal Fe₂O₃ concentration for enhancing AD performance; and (d) analyze the kinetics of methane production using appropriate mathematical models.

## 2. Materials and methods

The methodology is comprised of the following steps.

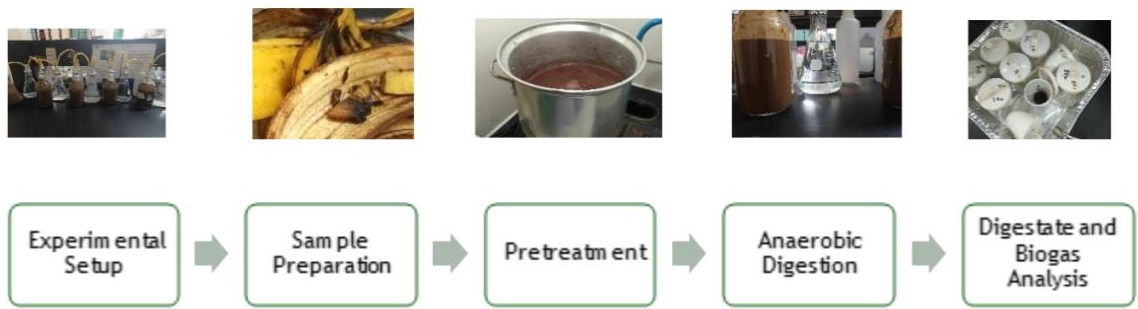

**Fig 1.** Summary of Methodology

### 2.1. Preparation of Feedstock

The substrates tested in the study were simulated SHWW which was collected from the Pasay City market and FW mixture consisting of 45% vegetable and fruit peels and 55% rice with used cooking oil will be acquired from household waste [3]. On the other hand, cow manure and distillery waste collected from Absolute Distillers were used as the inoculum for all the anaerobic digestion batches. The inoculum was activated by the addition of 30 mL of micronutrient solution (see Table 1) [4].

**Table 1.**
Summary of micronutrients

| Element | Micronutrients | Recommended concentration (mg/L) | Amount dissolved in 200 mL deionized water (g) |
|---|---|---|---|
| Fe | FeSO₄ 7H₂O | 0.3 | 0.015 |
| Cu | CuSO₄ 5H₂O | 0.6 | 0.03 |
| Zn | ZnSO₄ 7H₂O | 0.6 | 0.03 |
| Mg | MgCl₂ 6H₂O | 600 | 30 |
| Mn | MnCl₂ 4H₂O | 0.027 | 0.00135 |
| Co | CoCl₂ 6H₂O | 5 | 0.25 |

Meanwhile, powdered consortium enzymes manufactured by Infinita and bought from Gujarat, India, were used to improve the capacity of digesters to disintegrate complex organic matter by employing a synergistic approach [5]. Finally, the powdered ferric oxide, that promotes microbial activity required for effective anaerobic co-digestion, was sourced from Alysons' Chemical Enterprise in Manila.

*2.2. Pretreatment*

A ratio of 1:1 for FW and SHW was mixed for a final substrate solution of 560 mL per reactor [6]. The FW consisted of blended rice, vegetable waste, fruit waste, and used oil whereas simulated SHWW consisted of pig's blood with blended skin, fat, and liver. The SHWW and FW were mixed along with 62 mEq of calcium hydroxide per Liter and underwent chemical pretreatment at room temperature for 12 hours. Additionally, thermal pretreatment took place at 80°C for fifteen minutes to improve the degradation process. The samples were stored in a glass container and were subjected to hot water bathing for uniform heating. The mixture was cooled down at 30 °C before incorporating with consortium enzyme to transform complex high molecular weight compounds into methane [7]. On the other hand, the varying amounts of ferric oxide per run are summarized and presented in Table 2.

**Table 2.**
Experimental Design: 1:1 SHWW:FW

| Reactor | $Fe_2O_3$ mass (g/L) |
|---|---|
| 1 | - |
| 2 | 0.05 |
| 3 | 0.5 |
| 4 | 1 |
| 5 | 2 |

The anaerobic digestion tests were performed using 1000 mL bottles with 800 mL effective volume. The purging of nitrogen gas took place by using the glass tube submerged in the substrate-inoculum solution and was sealed with a rubber stopper [8]. A water displacement method was used to measure the amount of biogas generated from the anaerobic digesters. Moreover, the media bottles were connected to the displacement vessel which contains 1.5 M sodium hydroxide. The displaced NaOH in the third media bottle shall be recycled back to the displacement vessel. The reactor operating volume did not exceed 800 mL. Therefore, the anaerobic digestion of all samples was conducted for 30 days (Fig 2).

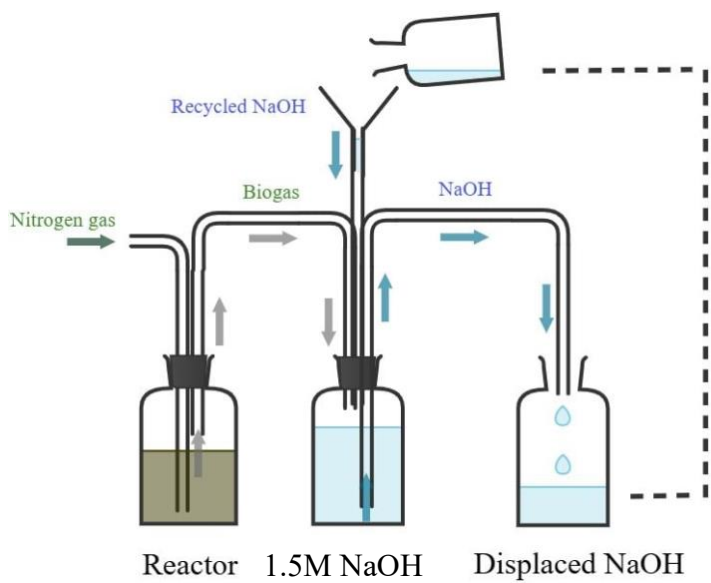

**Fig 2.** Experimental Set-up

*2.3. Analytical Methods*

All quality parameters were analyzed using the following methods found in Table 3. BOD and COD testing were conducted in Mach Union Laboratory in Las Pinas City whereas TS and VS were tested in YIC Research Laboratory at Mapua University. For the TS and VS testing, at least 10 grams of sample were collected from each reactor and were heated in the oven for 12 hours at 105 °C. This process was followed by the ignition of dried samples in the furnace at 550 °C for 2 hours.

**Table 3.**
Analytical Methods

| Parameter | Method (mg/L) |
| --- | --- |
| BOD | Azide Modification Dilution Technique |
| COD | Open Reflux Method |
| VS | Gravimetric Method |
| TS | Gravimetric Method |
| pH | D4927 Electrometric |

*2.4. Kinetics*

To understand the biogas production dynamics during anaerobic co-digestion, three well-established kinetic models were employed: the Modified Gompertz Model (Eq. 1, MGM), the

Modified Logistic Model (Eq. 2, MLM), and the Logistic Function (Eq. 3, LF) Model.

$$P(t) = K \exp(-\exp(-r(t - t_0))) \tag{1}$$

$$P(t) = \frac{K}{1 + \alpha \exp(-r(t - t_0))} \tag{2}$$

$$P(t) = \frac{K}{1 + \exp(-r(t - t_0))} \tag{3}$$

where $P(t)$ is the biogas produced at time, $t$, $K$ is the carrying capacity (maximum biogas production), $r$ is the growth rate, and $t_0$ is the time when the growth rate is highest.

### 3. Results and Discussion

*3.1. Feedstock Characterization*

Table 4 shows the physiochemical characteristics of the substrates utilized for anaerobic co-digestion of SHWW and FW. The TS and VS concentration of FW is depicted as high, which is considered a substantial organic load that is available for decomposition.

The optimal carbon-to-nitrogen (C/N) ratio for anaerobic digestion typically ranges between 20:1 and 30:1, ensuring a balance of carbon for energy and nitrogen for protein synthesis for microorganisms. FW has a high C/N ratio due to its higher carbon content while nitrogen-rich slaughterhouse wastewater has a lower ratio. The co-digestion of these waste streams can help achieve the overall C/N ratio, enhancing microbial activity, boosting biogas production, and maintaining process stability by adjusting their proportions.

*3.2. Effect of $Fe_2O_3$ on TS and VS*

Following the supplementation of the iron-based materials, it is important to evaluate the performance of the reactors on microbial activity, potential inhibitory effects, and the creation of toxic conditions through the resulting values of % reduction process efficiency. The data revealed a reduction efficiency for TS and VS concentrations from 52.98% - 58.82% and 53.85% - 69.47%.

**Table 4.**
Physiochemical characteristics of feedstocks

| Parameters | Substrates | | Inoculum | |
|---|---|---|---|---|
| | SHWW* | FW* | CM* | DWW* |
| pH | 7.1 | 5.91 | 7.25 | 3.80 |
| TS (%) | 3.5 | 13.88 | 18.5 | 7.69 |
| VS (%) | 3.2 | 41.97 | 68 | 6.49 |
| BOD (mg/L) | 2350 | 139.25 | - | 31250 |
| COD (mg/L) | 4502.5 | 189.26 | 24858 | 150840 |
| BOD/COD | 0.522 | 0.736 | - | 0.207 |
| C/N Ratio | 9.65 | 30.04 | 28.16 | 3.75 |
| TOC (%) | 32.62 | 19.57 | - | - |
| Total Nitrogen (%) | 3.38 | 2.11 | 1.53 | - |

*From literature

**Table 5.**
Comparison of TS and VS results

| Reactor | Initial TS (g/g) | Final TS (g/g) | % TS removal | Initial VS (g/g) | Final VS (g/g) | %VS removal | Biogas yield (mL) |
|---|---|---|---|---|---|---|---|
| R1 | 0.134 | 0.055 | 52.98 | 0.078 | 0.028 | 53.85 | 34.38819 |
| R2 | 0.114 | 0.049 | 58.07 | 0.072 | 0.022 | 68.06 | 46.94093 |
| R3 | 0.102 | 0.046 | 58.82 | 0.076 | 0.023 | 69.47 | 62.52498 |
| R4 | 0.116 | 0.049 | 54.31 | 0.072 | 0.023 | 63.06 | 41.26231 |
| R5 | 0.134 | 0.055 | 52.99 | 0.078 | 0.026 | 56.41 | 38.18565 |

It is observed from Table 5 that the percent TS and VS reduction are positively correlated to biogas production with R3 demonstrating the highest removal efficiency and biogas yield. Moreover, this implies a high conversion of organic matter to biogas at low amounts of ferric oxide.

In addition to that, the supplementation of $Fe_2O_3$ yields improved %TS and %VS reduction in comparison to the control (Fig 3). On the other hand, a higher removal efficacy of VS is indicated which shows evidence that significant amounts of volatile material are degraded by microbial communities present in the reactor [9].

Other than that, the BOD and COD decreased significantly in comparison to the initial data over 30 days (about 4 and a half weeks) of digestion (Table 6). The reduction in BOD shows that $Fe_2O_3$ aids in the elimination of biodegradable organic matter by coagulating and precipitating suspended solids. The diminishing in COD features the capacity of $Fe_2O_3$ to eliminate total organic matter through chemical precipitation and adsorption. The difference in BOD and COD for R1 and R3 indicates that the quality of substrates within the R3 has improved, which enhanced the biogas production for this digestor.

**Table 6.**
BOD and COD of Reactor 1 (Control) and Reactor 3

| Parameter | Before Anaerobic Co-digestion | | After Anaerobic Co-digestion | |
|---|---|---|---|---|
| BOD | Mixture of SHWW and FW | 143191 mg/L | Reactor 1 | 4131 mg/L |
| | | | Reactor 3 | 4049 mg/L |
| COD | | 323432 mg/L | Reactor 1 | 166981 mg/L |
| | | | Reactor 3 | 165038 mg/L |

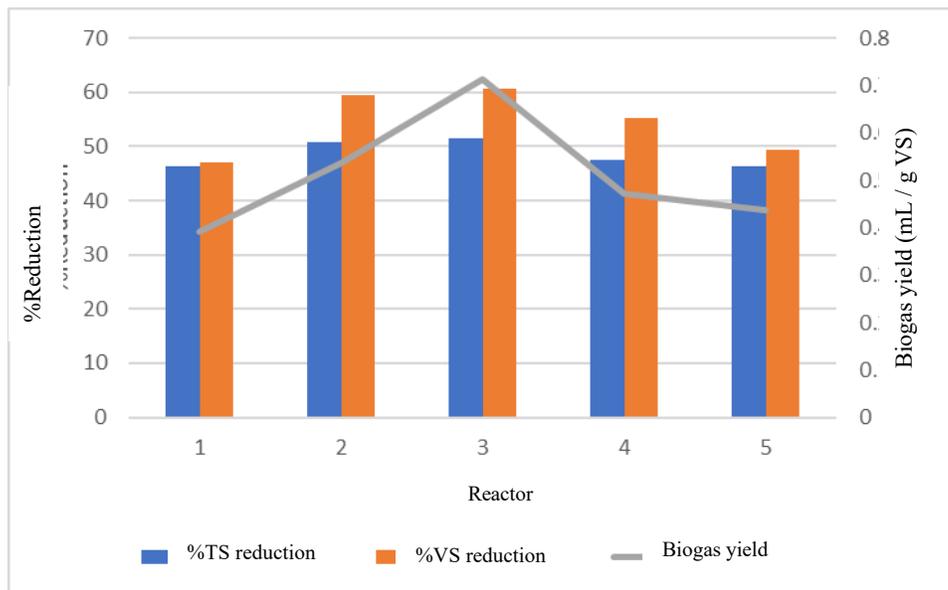

**Fig 3.** %TS and %VS Reduction

*3.3. pH Results*

The initial pH value of the substrate mixture was 6.78, which is within the optimal pH range for anaerobic digestion [10]. After 30 days of AcD, the final pH values of the reactors vary from 6.13 to 7.19. The reactors containing $Fe_2O_3$ were observed to have increased pH. Despite leaning into alkalinity, these values indicated a stable condition during the succeeding stages of digestion. The occurrence can be connected to the alkaline component generation and high buffering capability of the substrate [11]. Additionally, the substrates utilized in the study contain fats and high organic content that causes microorganisms to produce alkalinity, repressing the acidification of VFAs created. Meanwhile, the minimal decrease was seen in the control reactor (Fig 4), which is observed that acidification took place and hindered biogas production after the rapid valorization of organic matter or the accumulation of VFAs by the acidogenic bacteria [12].

Furthermore, the increase in pH was influenced by the ability of $Fe_2O_3$ to absorb inhibitory compounds that decreased the accumulation of organic acids concentration.

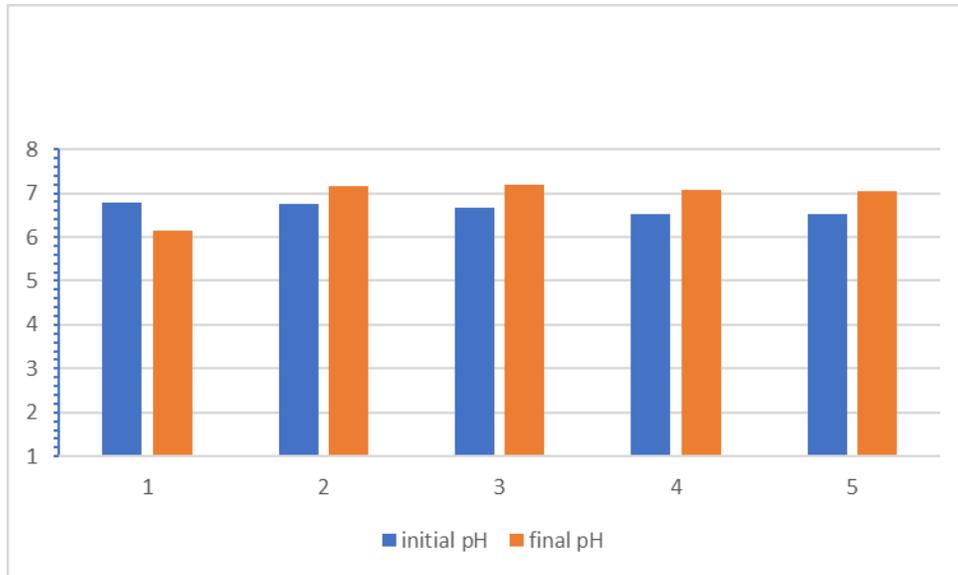

**Fig 4.** Initial and Final pH of the Reactors

*3.4. Biogas Production*

As specified in Table 2, each reactor has distinct concentrations of $Fe_2O_3$. Among all these, R3 produced significantly higher compared to the control reactor. While R5 exhibited an effective production, this was the first one to lose activity (Fig 5). Moreover, the outcome displayed the effect of different $Fe_2O_3$ concentrations, visualized in the figures below where the displaced water per grams of volatile solids (VS) added is plotted against the duration of anaerobic co-digestion. The amount of VS added is computed by dividing five grams of initial sample used to the acquired VS concentration of the sample. This graph depicts the effectivity of converting the organic matter present in reactor to biogas overtime.

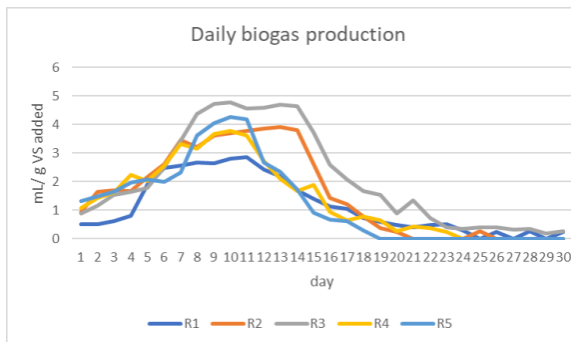 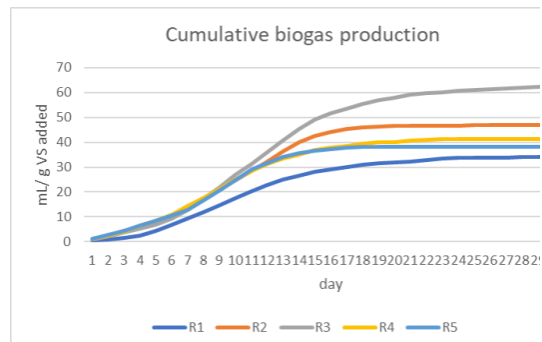

**Fig 5.** Daily Biogas Production        **Fig 6**. Cumulative Biogas Production

As observed, R1 (control) has produced the least biogas among the reactors but sustained production for the longest compared to reactors 2, 4, and 5. It peaked on the 16$^{th}$ day of anaerobic co-digestion with 3.8786 mL/g VS added of daily production. R3 peaked a day later achieving the highest daily yield of 4.6969 mL/g VS added, followed by R2 at 3.9030 mL/g VS. Meanwhile, R4 and R5 attained their maximum daily production days earlier, days 13 and 10 respectively. This may indicate that higher concentrations of $Fe_2O_3$ resulted in achieving their highest daily biogas production earlier which would also cause early exhaustion of organic materials for degradation.

In addition to that, the increase in biogas production was evident in the reactors with $Fe_2O_3$ compared to the control. Notably, R3 exhibited the largest amount of displaced water with 62.52 mL/g VS cumulative production and continuously produced biogas beyond the 30-day digestion time (Fig 6). This highlights the positive impact of ferric oxide in creating a conducive environment for microbial growth and biogas yield enhancement [13]. However, excessive amounts of ferric oxide were associated with suboptimal performance. Hence, incorporating an amount higher than 0.5 g in an 800 mL working volume may create a toxic environment for the microorganisms leading to early death phase in the digestion process.

To statistically validate these observations, T-testing was conducted to identify significant differences between the R1 and the rest of the reactors containing $Fe_2O_3$. It is observed from Table 7 that all reactors with iron-based material have a P-value less than 0.05, thus, these reactors are statistically significant.

**Table 7.**
T-Test Results

| Indication | | Reactors | R1 vs R2 | R1 vs R3 | R1 vs R4 | R1 vs R5 |
|---|---|---|---|---|---|---|
| Tails and Type | 1,2 | P-value | 2.49E-11 | 0.00057 | 0.02585 | 0.01585 |
| | 2,2 | | 4.98E-11 | 0.00115 | 0.05170 | 0.03170 |
| | 1,1 | | 0.0150 | 6.12E-10 | 4.01E-19 | 2.39E-12 |
| | 2,1 | | 0.0301 | 1.22E-09 | 8.03E-19 | 4.77E-12 |

Similar articles found on the Table 8 were studied to learn the difference of biogas production for mono-digestion and co-digestion of substrates with iron-based supplementation. In the present study, SHWW was usually co-digested due to its high nitrogen content which can be balanced by carbonaceous materials such as agricultural or food waste to stabilize the digestion process leading to a higher biogas yield and lower risks of ammonia inhibition [14]. Notwithstanding the evidence of the improvement of AcD, it is important to note that biogas production is still highly dependent on the type of substrate utilized and the specific conditions within the digester [15]. Furthermore, the addition of iron-based material like $Fe_2O_3$ would not only enhance the biogas yield but would also provide suitable environment for bacteria and microorganisms.

*3.5. Kinetics*

Kinetic parameters of the co-digestion process are essential to be studied. Therefore, it is necessary to model the anaerobic co-digestion process mathematically to predict the impact of organic matter mix ratios, the degree of organic loading, the substrate selection process, and production optimization [16]. The mathematical models used in this study are Modified Gompertz model (MGM), Modified Logistic model (MLM), and Logistic Function model (LF). MGM is frequently used for biological growth modelling such as microbial growth, cell growth, and animal growth where the growth rate decreases exponentially with time [17]. On the other hand, MLM is utilized in more intricate e ecological and biological models wherein growth rates shift from the standard logistic form permitting the model to consider different growth dynamics [18]. Unlike MGM and MLM which are usually engaged in modelling microbial growth, LF can be applied to various purposes and is also observed to have a symmetric curve around the inflection point that maximizes the growth rate [19].

**Table 8.**
Biogas production-related studies summary

| Substrate | Condition | Biogas Production | Reference |
|---|---|---|---|
| Slaughterhouse Waste + Food Waste | 30 days AD; mesophilic condition, uniform feeding of SHW and FW. | 4180 mL | [20] |
| Slaughterhouse Wastewater + Sewage sludge | 20 days AD; mesophilic condition, varying organic loading rates | 750 NL/kg VS | [21] |
| Slaughterhouse Wastewater | 30 days AD; Mesophilic condition, addition of iron nanoparticles | 600 NL/kg VS | [22] |
| Slaughterhouse Wastewater + Food Waste + Pig Manure | 15 days AD; Mesophilic condition, addition of $Fe_2O_3$ | 19.90 mL/g VS | [23] |

Hence, the study utilized the three models for calculating the substrates' biogas yield (A), maximum biogas yield rate (Rm), and lag phase duration (λ) using the experimental data on MATLAB.

Based on the results for $R^2$ presented in the Table 9, the experimental data is best fitted using the LF model. The resulting data from LF ($M_{calc}$) is compared with the experimental data ($M_{exp}$) using a 3D graph to illustrate the model's accuracy and precision. This implies that the said model can be a reliable tool for predicting biogas production and can be effectively used for an industrialized application of SHWW and FW co-digestion.

**Table 9.**
Mathematical model data

| Reactor | Model | A (mL) | Rm (mL/g VS.d) | λ (d) | Mcalc (mL/g VS) | Mexp (mL/g VS) | $R^2$ | Adjusted $R^2$ |
|---|---|---|---|---|---|---|---|---|
| R1 | MGM | 35.129 | 2.3889 | 4.8643 | 34.6943 | | 0.9957 | 0.9955 |
| | MRM | 35.8019 | 3.551 | 9.4361 | 32.8019 | 34.38819 | 0.9654 | 0.9636 |
| | LF* | 33.9636 | 2.4736 | 5.5075 | 33.9014 | | 0.9975 | 0.9974 |
| R2 | MGM | 52.0706 | 2.5291 | 3.4368 | 49.627 | | 0.9783 | 0.9772 |
| | MRM | 46.1611 | 4.8067 | 10.9179 | 46.1611 | 46.94093 | 0.9945 | 0.9943 |
| | LF* | 48.9232 | 2.7275 | 4.6258 | 48.4119 | | 0.9878 | 0.9872 |
| R3 | MGM | 66.8379 | 3.6355 | 5.2937 | 64.282 | | 0.9936 | 0.9933 |
| | MRM | 60.0369 | 5.575 | 10.844 | 60.0369 | 62.52498 | 0.9482 | 0.9454 |
| | LF* | 63.1353 | 3.864 | 6.2214 | 62.6202 | | 0.998 | 0.9979 |
| R4 | MGM | 43.2669 | 2.69652 | 2.98829 | 42.6564 | | 0.99081 | 0.99031 |
| | MRM | 41.4138 | 2.87683 | 4.16219 | 41.387 | 41.26231 | 0.99736 | 0.99722 |
| | LF* | 41.8686 | 2.78329 | 3.70147 | 41.7707 | | 0.99666 | 0.99648 |
| R5 | MGM | 38.5959 | 3.65746 | 3.08699 | 38.5593 | | 0.99710 | 0.99694 |
| | MRM | 37.3741 | 4.91544 | 6.24609 | 37.374 | 38.18565 | 0.96717 | 0.96539 |
| | LF* | 38.0498 | 3.66300 | 3.44565 | 38.0476 | | 0.99773 | 0.99761 |

On the contrary, the resulting lag phase duration (λ) ranges from 3.45 days to 6.22 days,

with R5 being the earliest and R3 the latest indicating its longer stabilization period. As per Wu et al [17], the introduction of foreign matter into the reactor promotes a longer lag phase as bacterial populations are adapting to new environmental conditions. Nevertheless, R5 having the highest amount of $Fe_2O_3$, suggested that higher concentrations of $Fe_2O_3$ stimulates direct interspecies electron transfer (DIET) and can potentially shorten the lag phase of the digestion process [19].

Meanwhile, it is observed from the results of maximum biogas yield (A) that the value of R1 was recorded at the lowest with 33.96 in contrast to R3 at 63.14. The trend in maximum biogas production aligns closely with the earlier observed parameters arranged in an increasing order respectively, R1<R2<R4<R5<R3. This further supports the observation that moderate $Fe_2O_3$ supplementation significantly enhances biogas production efficiency.

## 4. Conclusion

The co-digestion of SHWW and FW with iron-based supplementation demonstrates several key benefits and considerations. The high organic loading in FW, combined with the low nitrogen content in SHWW, necessitates balancing the C/N ratio for optimal anaerobic digestion. $Fe_2O_3$ enhances microbial activity, improves TS and VS reduction per cent and subsequently increases biogas production. Reactor 3, with moderate $Fe_2O_3$ levels, achieved the highest biogas yield and TS/VS reduction, indicating an effective balance of organic matter conversion. The observed increase in pH across reactors with $Fe_2O_3$ suggests improved digestion stability, while the control reactor exhibited less favorable conditions. Hence, the iron-based material acted as a catalyst by reducing inhibitory compounds, increasing biogas production and effluent quality, as well as amplifying DIET; this statement can be proved by the studies of Park, Aquino, and Fetra et al. [18, 21, 24].

Since excessive $Fe_2O_3$ concentrations led to diminished performance due to potential toxicity, the optimum concentration of $Fe_2O_3$ in the anerobic co-digestion of SHWW and FW is achieved at 0.5 grams for a working volume of 800 mL. Other than that, mathematical modeling using the Logistic Function model accurately predicted methane yield and process dynamics, showing its industrial applicability. Higher $Fe_2O_3$ concentrations can reduce the lag phase by enhancing electron transfer but may also deplete organic materials too soon. Thus, optimizing $Fe_2O_3$ levels is essential for maximizing biogas production and maintaining reactor stability. The study underscores $Fe_2O_3$'s potential to improve anaerobic digestion when used in appropriate amounts, emphasizing the need to balance additives and monitor their effects for optimal results.

For optimal performance, it is highly recommended that the reactors be effectively insulated to maintain a stable temperature during anaerobic digestion, as this process is susceptible to temperature fluctuations. Creating a mesophilic environment with a consistent temperature of approximately 37°C is crucial for promoting the growth of microorganisms involved in digestion.

Regular and thorough monitoring of the pH levels is also suggested, as it is essential for maintaining the process's stability. That said, VFA testing during and after anaerobic digestion is also encouraged to identify specific issues within the digester, allowing direct optimization of operation parameters.

Furthermore, it is recommended to explore and experiment with various techniques for effectively mixing $Fe_2O_3$ into the mixture, such as utilizing tea bags, to address the observed issues where the iron supplementation accumulates on the walls of the bottles or remains suspended at the top of the mixture. Other than that, the effects of various $Fe_2O_3$ particle sizes and concentrations ranging from 0.1-0.7g, which is around the obtained optimal ratio of 0.5g, should be assessed to identify the optimal amount that maximizes biogas production while avoiding toxicity.